\documentclass[preprint]{vgtc}               

\ifpdf
  \pdfoutput=1\relax                   
  \usepackage{graphicx}                
  \DeclareGraphicsExtensions{.pdf,.png,.jpg,.jpeg} 
\else
  \usepackage{graphicx}                
  \DeclareGraphicsExtensions{.eps}     
\fi%

\graphicspath{{figures/}{pictures/}{images/}{./}} 

\usepackage{microtype}                 
\PassOptionsToPackage{warn}{textcomp}  
\usepackage{textcomp}                  
\usepackage{mathptmx}                  
\usepackage{times}                     
\usepackage{cite}                      
\usepackage{tabu}                      
\usepackage{booktabs}                  

\usepackage{amsmath}
\usepackage{amsfonts}
\usepackage{subfig}

\onlineid{1171}

\vgtccategory{Research}

\vgtcinsertpkg

\title{Visualizing Similarity of Pathline Dynamics in 2D Flow Fields}

\author{Baldwin Nsonga\thanks{e-mail: nsonga@informatik.uni-leipzig.de}\\ %
        \scriptsize Leipzig University %
\and Gerik Scheuermann\thanks{e-mail: scheuermann@informatik.uni-leipzig.de}\\ %
        \scriptsize Leipzig University %
     }

\teaser{
  \centering
  \includegraphics[width=\linewidth]{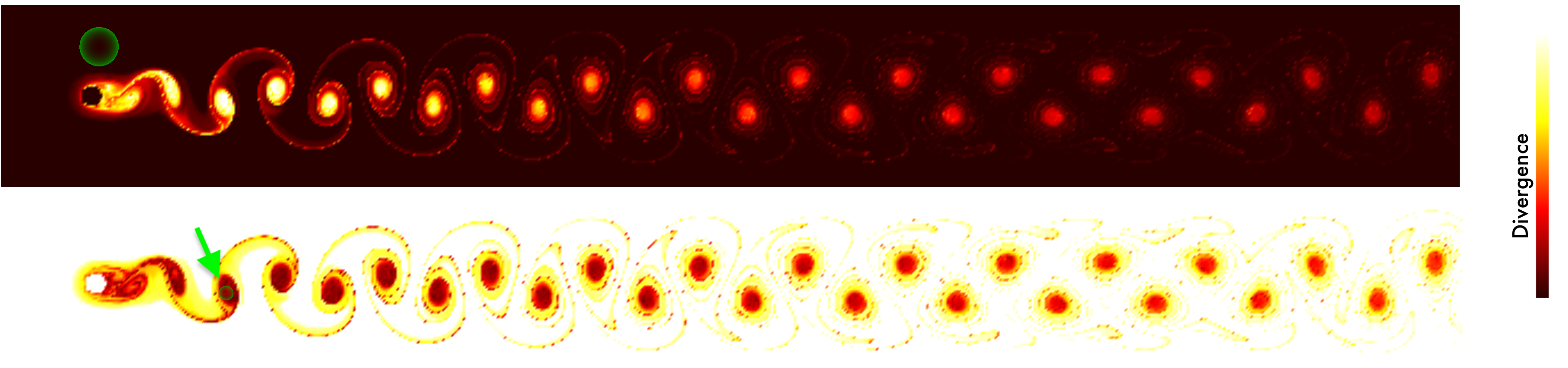}
  \caption{Visualisation of the divergence of the \textit{von Karman Vortex Street} data set with two distinct reference regions (green). The visualization was obtained using backward integration with the parameters: $t_0 = 15$, $\tau =15$ ,and $\Delta_t = 0.01$.}
  \label{fig:result:street}
}

\abstract{
Even though the analysis of unsteady 2D flow fields is challenging, fluid mechanics experts generally have an intuition on where in the simulation domain specific features are expected.
Using this intuition, showing similar regions enables the user to discover flow patterns within the simulation data.
When focusing on similarity, a solid mathematical framework for a specific flow pattern is not required.
We propose a technique that visualizes similar and dissimilar regions with respect to a region selected by the user.
Using infinitesimal strain theory, we capture the strain and rotation progression and therefore the dynamics of fluid parcels along pathlines, which we encode as distributions.
We then apply the Jensen–Shannon divergence to compute the (dis)similarity between pathline dynamics originating in a user-defined flow region and the pathline dynamics of the flow field.
We validate our method by applying it to two simulation datasets of two-dimensional unsteady flows.
Our results show that our approach is suitable for analyzing the similarity of time-dependent flow fields.%
} 

  \CCScatlist{
    \CCScatTwelve{Human-centered computing}{Visu\-al\-iza\-tion}{Visualization application domains}{Scientific visualization}
  }

\begin{document}


\firstsection{Introduction}

\maketitle

Researching flows is central to improving not only energy conversion processes but also understanding of natural phenomena such as ocean currents.
By highlighting meaningful structures, flow visualization can reveal underlying dynamics to further this understanding.
This is challenging, as fluid motion is a time-dependent, nonlinear, and possibly chaotic process.
Techniques concerned with feature detection or the computation of Largangian Coherent Structures were shown to be beneficial to understanding, although for some structures, such as vortices, a robust and objective definition is still a current object of research.

In this work, we propose a method that takes a user-defined region of interest and visualizes (dis)similarity.
The approach is based on pathlines for which we compute a strain/rotation progression using concepts from continuum mechanics. 
We construct dynamics distributions from the resulting progressions, which we use to compute similarity using the Jensen-Shannon divergence.
Our method does not require a clear definition of the feature that should be detected.
With the focus on dissimilarity, our method allows the user to select a base line, resulting in highlighting abnormal dynamics in relation to the chosen reference region.

\textbf{Notation: } 
We denote time-dependent velocity fields as $\mathbf{v}(\mathbf{x},t )$, where $\mathbf{x} \in D$ is a position in the flow domain $D \subset  \mathbb{R}^2$ and $t \in T$ is an instant in time within the time domain $T \subset \mathbb{R}$.
We denote pathlines by $p(\mathbf{x}_0, t_0; \tau )$.
A pathline describes the trajectory of a massless particle that is advected by the underlying velocity field.
It is a solution of the differential equation $\frac{d}{ d t }\mathbf{x}( t ) = \mathbf{v}(\mathbf{x}(t), t)$, with $\mathbf{x}(t_0) = \mathbf{x}_0$. The parameter $\tau$ describes the integration time.

\section{Related Work}

A common approach to visualize flows is to focus on specific flow patterns with spatial and temporal extent, that is, features. 
Features (e.g. vortices \cite{ hunt1988qcrit, jeong1995lambda2,Okubo:1970:okubo-weiss-criterion ,Chong:1990:vortex}, shock waves \cite{wu2013shockwaves}, splats \cite{Nsonga:2019:splats}) allow for a purposive investigation of flow behavior, as they are associated with specific physical properties of the flow.
Feature detection methods, for example, the $\lambda_2$ \cite{jeong1995lambda2} or $Q$-criterion \cite{hunt1988qcrit}, have drawbacks, as the results depend on the chosen reference frame.
A body of research is concerned with objective flow characterization \cite{Haller:2005:Mz,  Bhatia2014internal_reference, Gunther2017generic_objective, Wiebel2005localized_flow, Wiebel2007localized_flow_computation} or \cite{Haller:2021:single_trajectory} for sparse trajectory data.

Instead of investigating specific features, the vector field topology \cite{Helman1989Flow_topology, Helman1991Flow_topology} was shown to be suitable for analyzing the underlying structure of flows, where the so-called topological skeleton is calculated and depicted.
The topological skeleton consists of critical points, i.e. points in space in which the velocity vanishes, and separatrices seperating regions of the flow domain.
Streamlines seeded within regions constructed by the separatices share the same source and sink.
In their work, Günther and Rojo \cite{gunther2021introduction} provide a general overview of this class of techniques. 
Although vector field topology is only applicable to steady flow fields, Bujack et al. \cite{Bujack:2020:topology} provide a survey of techniques exploring time-dependent flow topology.
The time-dependent analogy for vector field topology are Lagrangian Coherent Structures (LCS)\cite{Haller:2015:lagrangian}.
The Finite-Time Lyapunov Exponent (FTLE) field proposed and discussed by Haller \cite{Haller:2000:FTLE} is the most common technique used for LCS visualization.

Although we will utilize ideas developed in the context of LCS, our goal is not a direct detection of a specific flow pattern.
Our method should extend existing techniques in such a way that, for example, a similarity field with respect to a detected feature can be computed fast.

In the context of similarity in 2D vector fields, Schlemmer et al. \cite{Schlemmer:2007:moments1, Schlemmer:2007:moments2}  and
Bujack et al. \cite{bujack2015moment} proposed methods applying moment invariants for pattern matching.
After visual selection, Bujack et al. \cite{bujack2015moment} use statistical moments and normalization as rotation, translation, and scale invariant descriptors. 
In the following work \cite{Bujack:2015:clustering_moments}, they used their technique to cluster similar flow patterns without predefined descriptors.

\section{Background}
\label{sec:background:ftle}


\subsection{Localized FTLE}
\label{sec:localized_ftle}
Localized FTLE was introduced by Kasten et al. \cite{Kasten:2009:localized} and is an alternative to computing FTLE \cite{Haller:2000:FTLE} using flow maps.
Flow maps are constructed by advecting particles in the flow forward (or backward) in time.
The flow map maps the initial positions of the tracer particles to their end positions after a specified integration time.
The map gradient then describes the deformation gradient used for computing the right Cauchy-Green deformation tensor.
Its spectral norm normalized by the integration time results in the FTLE field.
In contrast, \textit{localized FTLE} is based on the local deviation of the neighborhood along a pathline and allows the pathlines to be processed individually.

With the velocity field $\mathbf{v}(\mathbf{x},t )$ and a pathline $ p(\mathbf{x}_0, t_0; \tau )$, the deviation of infinitely close particles is governed by the spatial gradient of the velocity field along the pathline.
We denote an infinitely close perturbed particle as $\mathbf{y} := \mathbf{x}_0 + \sigma(t)$, with the time-varying perturbation vector $\sigma(t) \in \mathbb{R}^3$.
At $t_0$,  $\|\sigma(0)\|$ is close to zero.
Kasten et al. then approximated the temporal evolution of the deviation $\sigma (t)$, by considering very short advection times.
The solution of the differential equation for short advection times is then $\sigma(t) = \mbox{exp}(\nabla_0 t) \sigma(0)$, where $\nabla_0 = \nabla \mathbf{v}(p(0), t_0)$

Kasten et al. \cite{Kasten:2009:localized} applied this approximation for short time intervals along the pathline $p(\mathbf{x}_0, t_0; \tau )$.
Note that the samples are equidistant in time.
They obtain the following description for the cumulative deviation along a pathline $\sigma(\tau) =\Psi_{t_0}^{t_0+\tau} (\mathbf{x}_0)  \sigma(0)$,
where the matrix:
\begin{equation}
\label{eq:psi}
\Psi_{t_0}^{t_0+\tau} (\mathbf{x}_0) := \prod_{i = N-1 }^0 \mbox{exp}( \nabla \mathbf{v}(p( i \Delta_t ), i\Delta_t)  \Delta_t )
\end{equation}
maps the neighborhood of the point $\mathbf{x_0}$ to the deviation after advection over the advection time $\tau$.
$N$ is the number of discrete time steps, and $\Delta_t$ is the (small) temporal sample distance.
The integration time satisfies $\tau = \Delta_t \cdot N$.
$\Psi_{t_0}^{t_0+\tau} (\mathbf{x}_0)$ can be treated analogously to the flow map gradient.
Although we will not apply localized FTLE directly, we will utilize the idea of computing the deformation gradient along pathlines by accumulating small deviations.

\subsection{Jensen-Shannon Divergence}
The Jensen-Shannon divergence (JSD) \cite{Lin:1991:JSD} is an information-theoretic divergence measure based on Jensen's inequality \cite{jensen:1906:inequality7} and the Shannon entropy \cite{Shannon:1948:entropy}.
It is similar to the common Kullback-Leibler divergence \cite{Kullback1951}, which measures the divergence between statistical populations.

The Kullback-Leibler divergence is defined as follows:
\begin{equation}
        KL(\xi_1 | \xi_2) = \sum_{x \in X} \xi_1(x) \mbox{log}  \frac{\xi_1(x)}{\xi_2(x)},
\end{equation}
where $\xi_1$ and $\xi_2$ are discrete probability distributions defined in the same probability space $X$.
The Kullback-Leibler divergence is always larger than zero.
As $KL(\xi_1 | \xi_2) \neq KL( \xi_2| \xi_1)$ applies, this measure is not symmetric.
Note that it can be made symmetric by defining the measure $J = KL(\xi_1 | \xi_2)  + KL(\xi_2 | \xi_1) $.
The most significant drawback for our purposes is that the Kullback-Leibler divergence requires absolute continuity: $\xi_2(x) = 0 \implies \xi_1(x) = 0$.
In our approach, absolute continuity is not guaranteed.

In our method, we apply the closely related Jensen-Shannon divergence:
\begin{equation}
    JSD( \xi_1 , \xi_2 ) = \frac{1}{2} KL( \xi_1 | m  ) + \frac{1}{2} KL( \xi_2 | m ),
\end{equation}
where $m = \frac{1}{2} \xi_1 + \frac{1}{2} \xi_2$.
The JSD is a symmetric divergence measure that does not require absolute continuity, as $m(x) = 0 \implies \xi_1(x) = 0 \land \xi_2(x) = 0$ applies by definition.




\section{Methodology}
\label{sec:method}

\begin{figure*}[!t]
\centering
\subfloat[FTLE]{\includegraphics[trim={0 2cm 0 1.5cm},clip, width=0.12\linewidth]{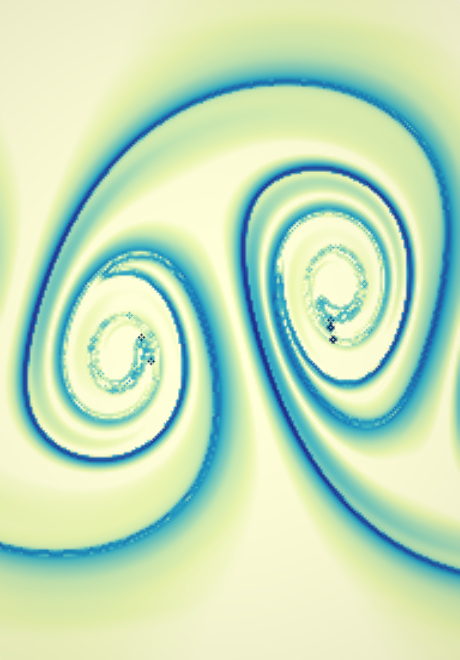}%
}
\hfill
\subfloat[$\Delta_t = 0.01$]{\includegraphics[trim={0 2cm 0 1.5cm},clip,width=0.12\linewidth]{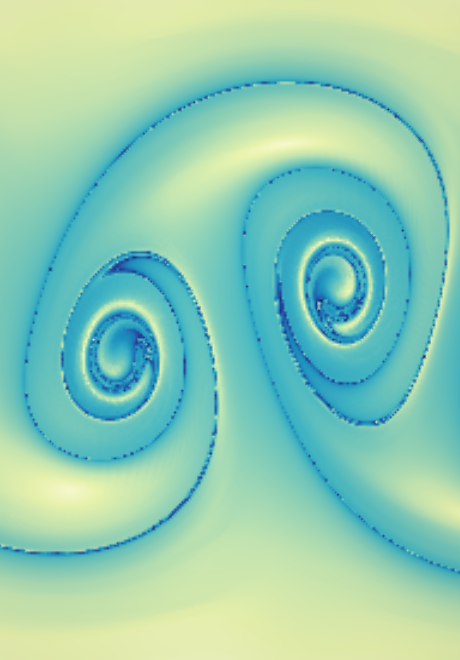}%
}
\hfill
\subfloat[$\Delta_t = 0.05$]{\includegraphics[trim={0 2cm 0 1.5cm},clip,width=0.12\linewidth]{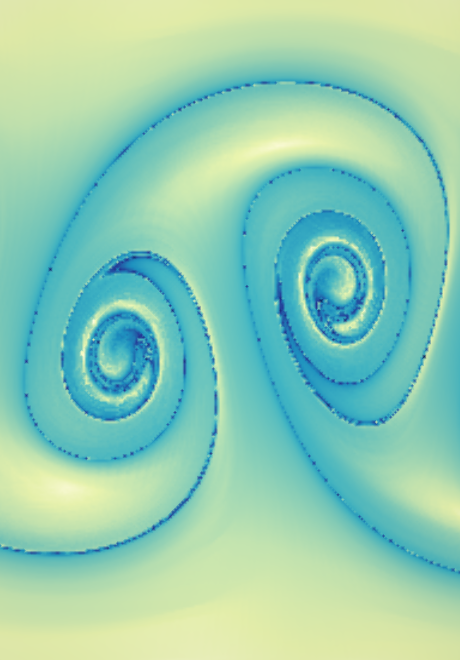}%
}
\hfill
\subfloat[$\Delta_t = 0.1$]{\includegraphics[trim={0 2cm 0 1.5cm},clip,width=0.12\linewidth]{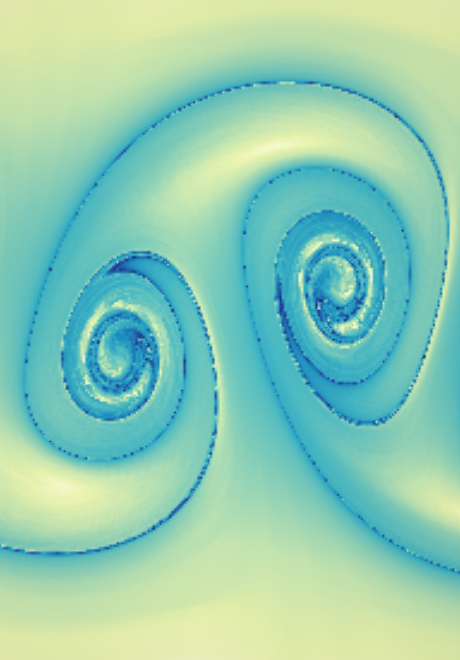}%
}
\hfill
\subfloat[$\Delta_t = 0.5$]{\includegraphics[trim={0 2cm 0 1.5cm},clip,width=0.12\linewidth]{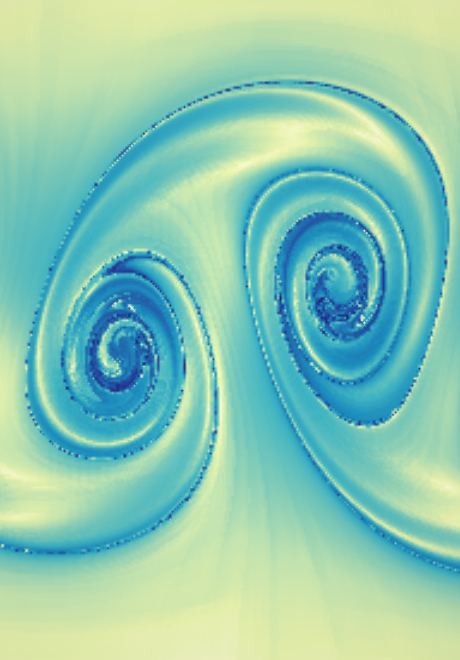}%
}
\hfill
\subfloat[$\Delta_t = 1$]{\includegraphics[trim={0 2cm 0 1.5cm},clip,width=0.12\linewidth]{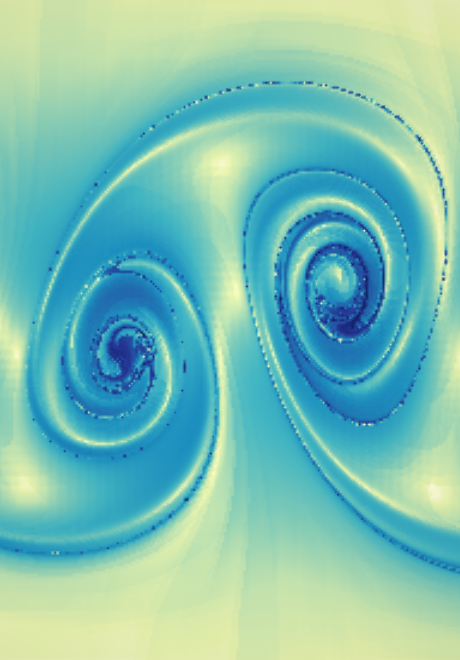}%
}
\hfill
\subfloat[$\Delta_t = 2$]{\includegraphics[trim={0 2cm 0 1.5cm},clip,width=0.12\linewidth]{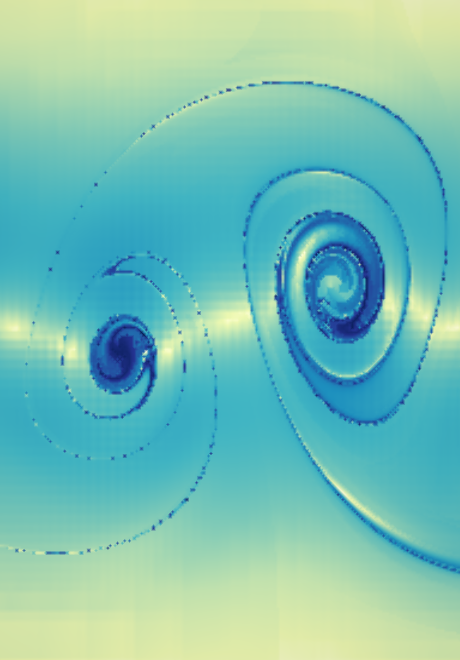}%
}
\caption{Visualization of the FTLE field (a) using reconstruction from strain progression for varying sample distances (b,c,d,e,f,g). The visualization was carried out on the \textit{von Karman Vortex Street} data set (cf. \autoref{sec:datasets}) with subdivision using backward integration using parameters $t_0 = 15$ and $\tau =15$. }
\label{fig:error}
\end{figure*}

Our method consists of two steps. 
In the first step, we construct a \textit{dynamics distribution}, which represents the stretching and rotation that a fluid parcel undergoes during a specific timeframe along its trajectory.
In the second step, taking a user-defined reference region, we compute the similarity of the dynamics distributions to the \textit{reference distribution} using the Jensen-Shannon divergence.

In the following, we first describe how we utilize ideas from localized FTLE and infinitesimal strain theory to compute a set of transformations that describe the dynamics along a pathline.
We then discuss how we apply these distributions for a fast computation of the similarity of pathline dynamics with respect to a reference distribution.

\subsection{Dynamics Distribution}
The finite-strain tensor is a common tensor in continuum mechanics and closely related to the right Cauchy-Green tensor.
It is defined as $E = \frac{1}{2} [F^T F   - I ] = \frac{1}{2} [C   - I ]$,
where $F \in \mathbb{R}^{2 \times 2}$ is the deformation gradient, $C \in \mathbb{R}^{2 \times 2}$ is the right Cauchy-Green tensor, and $I \in \mathbb{R}^{2 \times 2}$ is the identity.
Analogously to the Cauchy-Green tensor, it describes the deformation of a fluid volume from a reference configuration and captures nonlinear dynamics.

Note that the product integral described in \autoref{eq:psi} computes the product of matrices.
As matrix multiplication is not commutative, the gradient for short advection times must be applied in the correct order.
By directly constructing and utilizing a distribution from the deformation gradients for short integration times that contribute to $\Psi$, we lose information on the correct order. 
Our goal is to construct a distribution based on (short) deformation gradients without order dependence, and thus without matrix multiplication.
In order to describe the deformation dynamics as a distribution, we first require a description of the nonlinear strain tensor using summation, which is commutative.

After computing the pathlines (using the Dormand-Prince method \cite{Dormand:1980:RK}), we construct an infinitesimal strain tensor by applying the localized concept for short advection times.
The deformation gradient for a short advection time is described and approximated by omitting the higher order terms of the Taylor series:
$ F_{\Delta}  =   \mbox{exp}(\nabla_0 \Delta_t)) \approx I + \nabla_0 \Delta_t$,
where $\Delta_t \in \mathbb{R}$ is the (small) advection time, and $\nabla_0$ is the spatial gradient of the velocity field at the corresponding position in space and time.

We can now calculate the finite strain for short advection times by inserting $F_{\Delta}$ into the equation of $E$ and rearranging the equation to: $E_{\Delta} = \frac{1}{2}[ \Delta_t\nabla_0 + \Delta_t\nabla_0^T + \Delta_t^2 \nabla_0^T \nabla_0]$.
We choose the advection time to be very short ($\Delta_t \ll 1 $), so that $\|\Delta_t \nabla_0  \|$ is also small.
We can then omit $\Delta_t^2 \nabla_0^T \nabla_0$, as $\Delta_t^2 \ll \Delta_t$.
This results in the so-called infinitesimal strain tensor, a common tensor from continuum mechanics: $\mathbf{\epsilon} = \frac{1}{2}[ \Delta_t\nabla_0 + \Delta_t\nabla_0^T] \approx E_{\Delta}$.

We now briefly show that the sum of infinitesimal strains along a pathline is a viable approximation for the finite strain tensor.
Two consecutive transformations $\mbox{exp}(\nabla_0 \Delta_t))$ and $\mbox{exp}(\nabla_1 \Delta_t))$  result in the deformation gradient $F_2 = \mbox{exp}(\nabla_1 \Delta_t))  \mbox{exp}(\nabla_0  \Delta_t))$.
When inserting $F_2$ into the equation for the finite-strain tensor, substituting for the approximation, and omitting the matrix products as shown in the equation for $E_\Delta$ and $\mathbf{\epsilon}$, we obtain the following equation:
\begin{equation}
   E_2 \approx \frac{1}{2}[ \Delta_t\nabla_0 + \Delta_t\nabla_1 + \Delta_t\nabla_0^T + \Delta_t\nabla_1^T ] =  \mathbf{\epsilon}_0 +  \mathbf{\epsilon}_1.
\end{equation}

Note that this only holds if $\|\Delta_t \nabla_0  \| \ll 1$. 
This construction is analogous to the construction of more consecutive transformations.
This results in $\sum_{i=0}^{N-1} \mathbf{\epsilon}_i( \mathbf{x}_0, t_0; \Delta_t)$, which can be interpreted as a strain progression.
This progression is the basis for our deformation dynamics.
Note that the approximation error can accumulate for many terms. 
Here, we study the error introduced by our approach visually.
\autoref{fig:error} shows the resulting FTLE field (far left) and the calculation of the principal stretch using the sum of the infinitesimal strain to approximate the finite stain.
It is clear that an error is introduced, yet we found it to be tolerable for our approach, as the figure suggests a strong correlation between the FTLE values and the approximation sum.

In addition to the infinitesimal strain $\epsilon$, we also consider the infinitesimal rotation tensor: $\mathbf{\omega} = \frac{1}{2}[ \Delta_t\nabla_0 - \Delta_t\nabla_0^T]$.

We found that the invariant of the infinitesimal strain tensor $\epsilon$ and the rotation tensor $\omega$ are viable for the construction of the transformation distribution.
Note that $\mbox{tr}(\omega) = 0$ and for incompressible fluids $\mbox{tr}(\epsilon) = 0$ apply.
Here, we chose $\alpha = \mbox{det}(\epsilon) \in \mathbb{R}$ and $\beta = \mbox{det}(\omega) \in \mathbb{R}$.
$\alpha$ can be interpreted as the infinitesimal squared principal stretch of the fluid parcel, which is an objective measure.
$\beta$ can be interpreted as the infinitesimal vorticity magnitude.
As vorticity is not objective, we assume the resulting distribution to be not objective.

We now construct the progression set by first computing $\{ \alpha_0, \alpha_1, \ldots, \alpha_{N-1} \}$ and $\{ \beta_0, \beta_1, \ldots, \beta_{N-1} \}$ along a pathline using the defined sample distance $\Delta_t$.
$N = \tau / \Delta_t$ applies, where $\tau$ is the integration time.
We then compute a histogram $\xi_p$ by first constructing a histogram for the set $\alpha$ and the set $\beta$, respectively.
The two one-dimensional histograms with $n$ bins are then concatenated to form a one-dimensional histogram with $2n$ bins.
Note that constructing a two-dimensional histogram would imply that the sets of invariants are independent, which is not the case, as they both depend on $\nabla_0 \Delta_t$.
We normalize $\xi_p$ so that the sum of all bins equals one.

\begin{figure}[!tb]
\centering
\subfloat[n $= 10$ ]{\includegraphics[width=0.49\columnwidth]{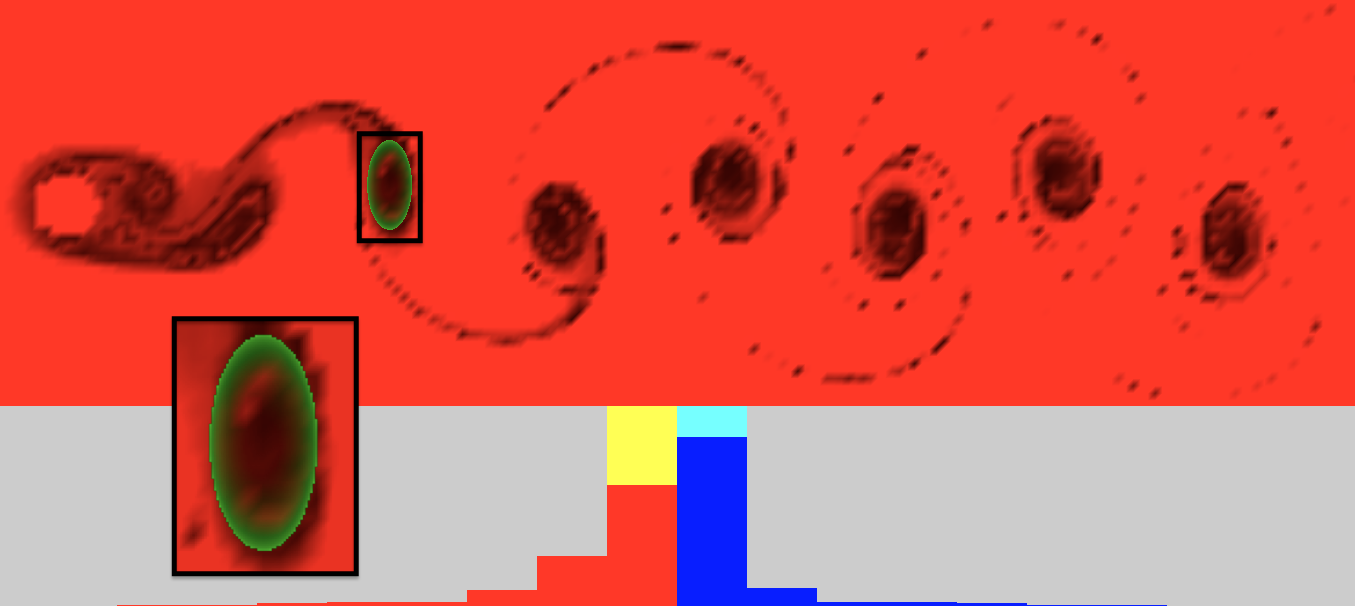}%
}
\hfill
\subfloat[$n = \sqrt{ N } = 38$ ]{\includegraphics[width=0.49\columnwidth]{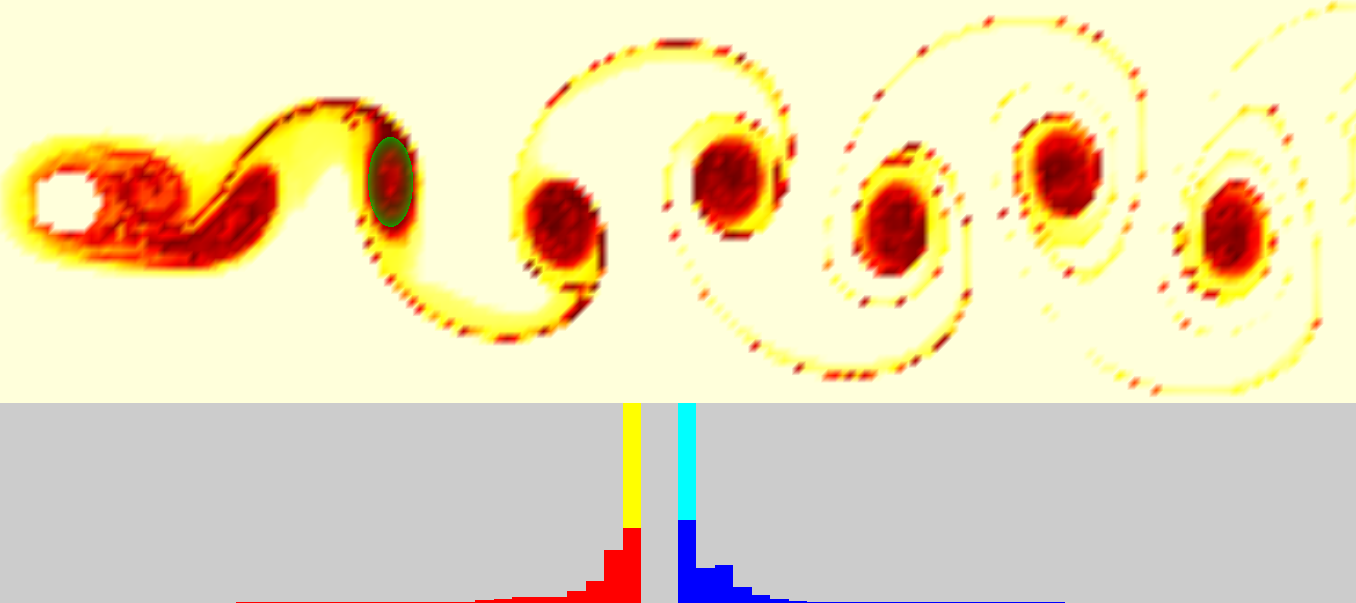}%
}
\hfill
\subfloat[$n = 100$ ]{\includegraphics[width=0.49\columnwidth]{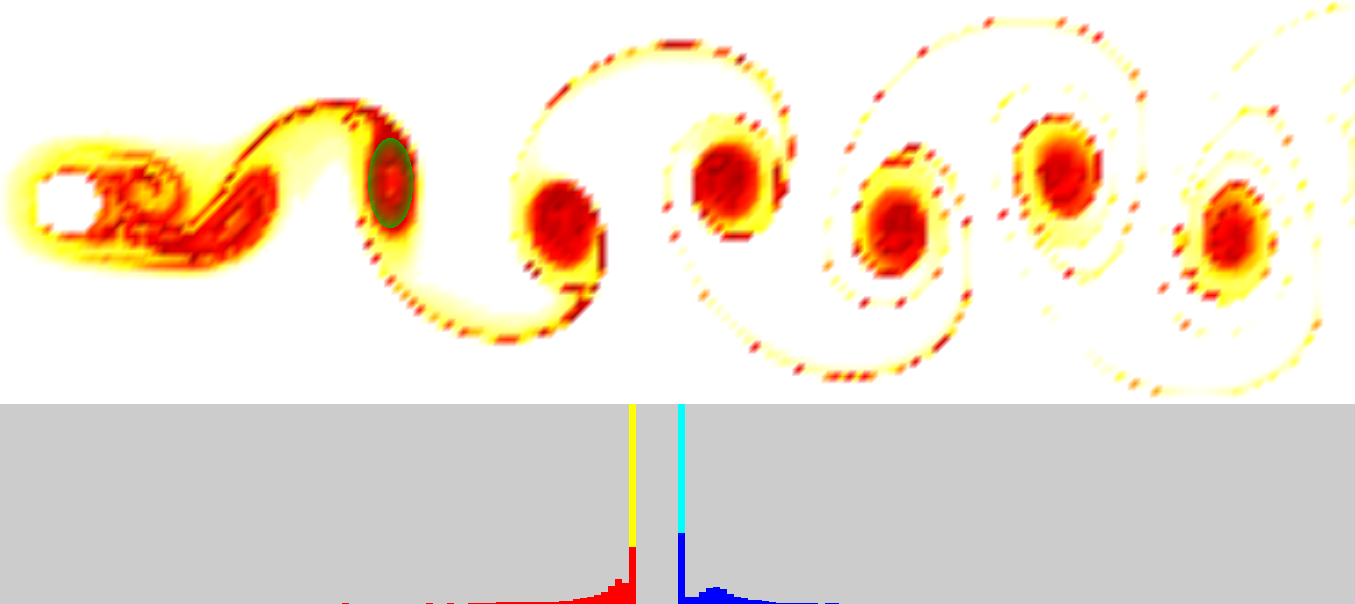}%
}
\hfill
\subfloat[$n = 200$ ]{\includegraphics[width=0.49\columnwidth]{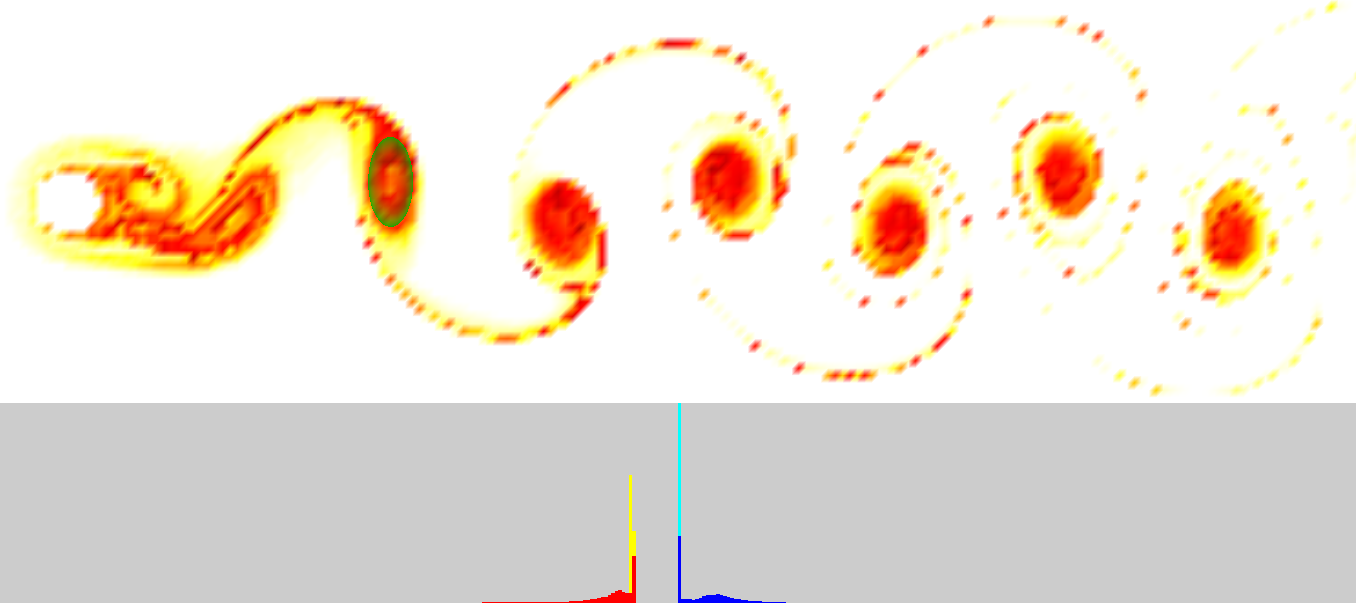}%
}
\caption{ Visualization of the divergence and the corresponding distributions of the \textit{von Karman Vortex Street} using an elliptic reference region. The visualization was obtained by integrating backward with parameters $t_0 = 15$, $\tau =15$, and $\Delta_t = 0.01$ with a varying number of bins $n$. The distributions of $\alpha$ (red), $\beta$(blue) for the reference configuration and the most dissimilar pathline (yellow, cyan, respectively) are shown.}
\label{fig:bins}
\end{figure}

\subsection{Similarity Visualization} 
Our method is based on the visualization of the similarity of pathline dynamics using a reference deformation distribution and the deformation distribution computed earlier.
For visualization, we first compute the reference deformation distribution $\xi_R$.
The user defines an area $R \subseteq \mathbb{R}^2$ for the reference distribution. 
Here, we use circular regions, but any region is applicable.
Now, we construct a distribution for every pathline, where $\mathbf{x}_0 \in R$ by adding the bins of the corresponding pathlines.
After normalizing by the number of elements, we obtain the reference distribution $\xi_R$.

We can now calculate the similarity by computing the Jensen-Shannon divergence $JSD(\xi_p, \xi_R) $.
Small divergence values indicate strong similarity, where a value of zero implies that $\xi_R$ and $\xi_p$ are the same distribution.
Here, we normalize the resulting field by the upper bound, which is $\ln(2)$, when using the natural logarithm.
The divergence is then displayed by color mapping.

\subsection{Parameters}
\label{sec:results}

\textbf{Amount of Bins:} 
The amount of bins determines the quality of the dynamics distribution. 
Choosing a value too low leads to a greater variance of the values $\alpha$ and $\beta$ within a bin. 
As a result, the algorithm can overestimate the similarity between distributions, leading to overall low divergence values.
High values increase the computation time and produce a large amount of empty bins, especially if the number of samples is low.
We found that the common rule of thumb $n = \sqrt{ N }$ produces viable results and a solid starting point for exploration.
$N$ can be directly computed from the chosen integration time and sample distance (cf. \autoref{fig:bins}).

\begin{figure}[!tb]
\centering
\hfill
\subfloat[$\tau = 1$ ]{\includegraphics[trim={0 0.5cm 0 0.5cm},clip,width=0.48\columnwidth]{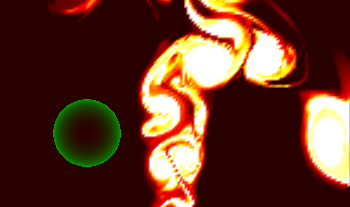}%
}
\hfill
\subfloat[$\tau = 10$ ]{\includegraphics[trim={0 0.5cm 0 0.5cm},clip,width=0.48\columnwidth]{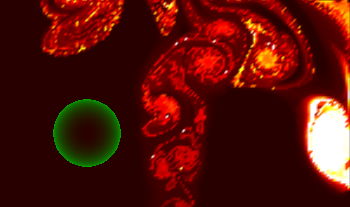}%
}
\hfill
\subfloat[$\Delta_t = 0.5$ ]{\includegraphics[trim={0 0.5cm 0 0.5cm},clip,width=0.48\columnwidth]{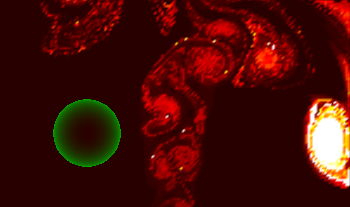}%
}
\hfill
\subfloat[$\Delta_t = 0.01$ ]{\includegraphics[trim={0 0.5cm 0 0.5cm},clip,width=0.48\columnwidth]{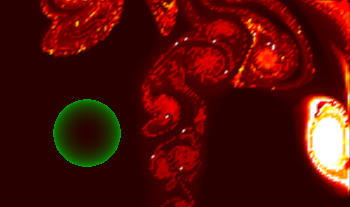}%
}
\caption{ The influence of the parameters shown on the buoyancy dataset. The non-varying parameters are set to $\Delta_t = 0.01$ and $\tau = 20$ with a bin number of $100$. The integration was performed backward in time starting at $t_0 = 13$.}
\label{fig:sample_distance}
\end{figure}

\textbf{Sample Distance $\Delta_t$:} 
The sampling distance determines the resolution of our pathline approximation and the quality of the infinitesimal strain tensor.
It is straightforward to assume that the parameter should be chosen as small as possible.
Increasing the resolution and thus decreasing $\Delta_t$ leads to a longer computation time for the dynamics distribution.
A high value of $\Delta_t$ can lead to artifacts due to the omission of higher-order terms that have a greater impact for higher values of $\Delta_t$ (cf. \autoref{fig:sample_distance}e,f).

\textbf{Integration Time $\tau$:} 
From the construction of the dynamics distribution, it is apparent that a decrease in integration times leads to a decrease in variance.
This may lead to a more pronounced divergence.
It is desirable to study long pathlines as it reveals more details in the visualized flow structures associated with dissimilarity (cf. \autoref{fig:sample_distance}c,d).

\section{Results}
\label{sec:results}

\begin{figure}[!tb]
	\centering
	\includegraphics[trim={0 5cm 0 6cm},clip,width=\columnwidth]{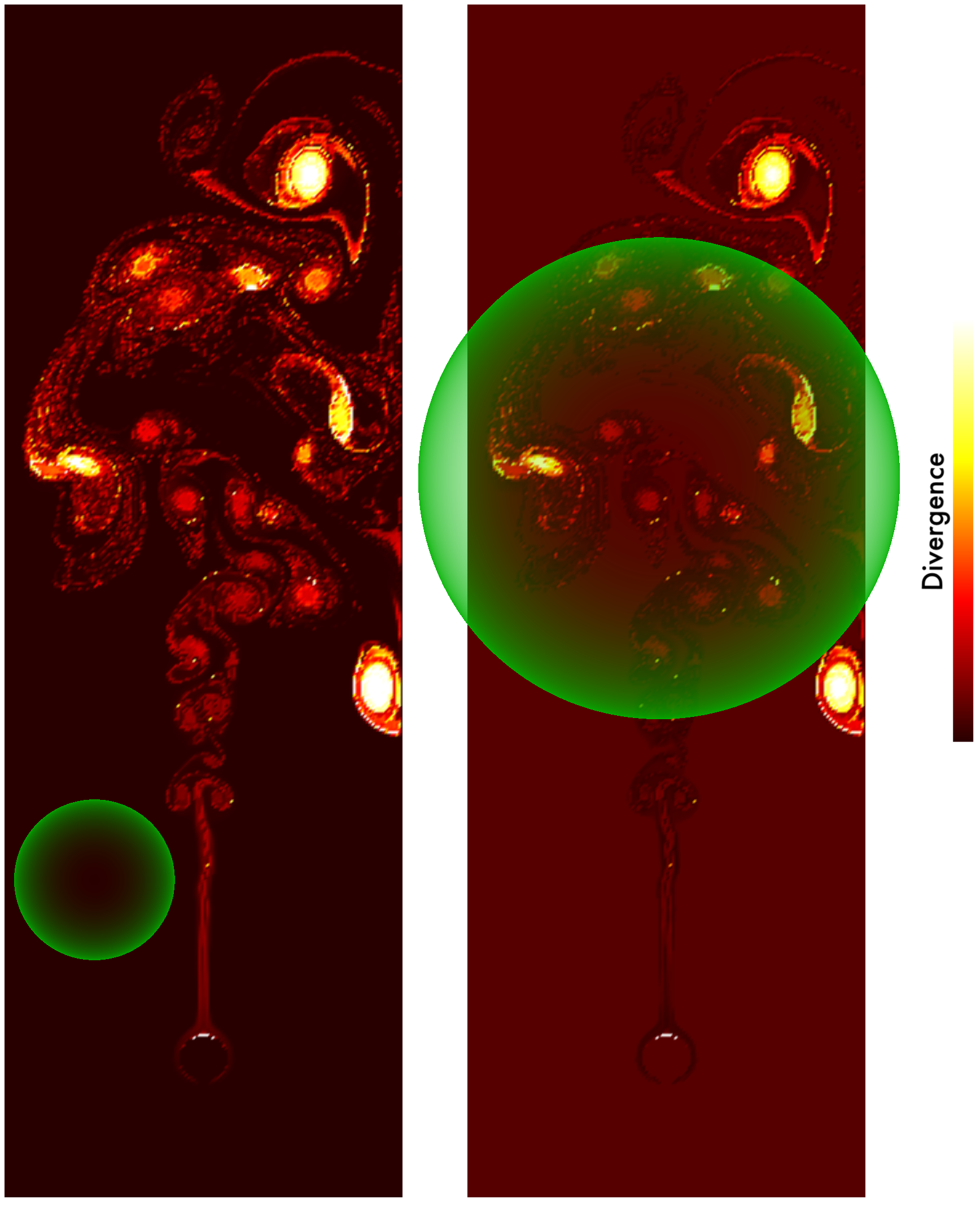}
	\caption{Visualisation of the divergence with two distinct reference regions (green). The visualization was obtained using backward in time integration with the parameters: $t_0 = 13$, $\tau =20$ ,and $\Delta_t = 0.01$.}
	\label{fig:result:heated}
\end{figure}

\subsection{Datasets}
\label{sec:datasets}

We explore two data sets that describe a time-dependent 2D flow.
They are provided by the Computer Graphics Laboratory of ETH Zürich and were used in the work of Günther et al. \cite{Gunther2017generic_objective}.
The data sets were generated using the Gerris flow solver \cite{gerrisflowsolver}.
The first data set represents the common flow pattern of a \textit{von Karman Vortex Street}, where the flow is disturbed by a cylinder.
The second data set describes a flow that develops due to buoyancy effects caused by a heated cylinder. 
The vortex street spans $[-0.5, 7.5] \times [-0.5, 0.5]$ with a grid resolution of $640 \times 80$.
It consists of $1501$ time steps in the range of $[0,15]$.
The heated cylinder data set spans the domain $[-0.5, 0.5] \times [-0.5, 2.5]$ with a grid resolution of $150 \times 450$.
It consists of $2001$ time steps in the range of $[0,20]$.

\subsection{Performance}

The following computation times were obtained on a Linux system with Ubuntu 20.04 LTS, 32GB of RAM, and an Intel(R) Xeon(R) CPU E5-2630 v3 @ 2.40GHz chip (2 physical processors and 32 threads).
This construction of the dynamics distribution is the most costly, as it requires numeric integration for every point.
This is only required once for a specified timestep, integration time, and sample distance.
Note that we store the dynamics sets $\alpha$ and $\beta$ for a fast recalculation of the reference distribution.
In a grid with $M$ vertices and input parameters $\tau$ and $\Delta_t$, this results in values $M \tau / \Delta_t$ that must be stored in memory.
This results in a large but manageable memory footprint for 2D flows on modern workstations.
Using the rule of thumb mentioned above, the distributions have a memory footprint of $M \sqrt{\tau \Delta_t}$. 
For the buoyancy data set with the parameters specified in \autoref{sec:results}, we measured a computation time for the construction of the dynamic distribution of approximately 13 seconds.
The computation of the similarity visualization takes approximately 1 second for this case and does not require additional memory.
As the integration and computation of $\alpha$ and $\beta$ are independent of each other, we parallelized these computations using OpenMP.

\subsection{Use Cases}
\label{sec:results:use_case}

The resulting visualization is shown in \autoref{fig:result:street} and \autoref{fig:result:heated}.
The reference region in \autoref{fig:result:street} (top) and \autoref{fig:result:heated}(left) corresponds to the use case in which the user investigates the dissimilarity (high divergence) from a known base line.
This reveals abnormal behavior, which is in this case associated with vortex shedding or, in the case of the heated cylinder, turbulent motion.
\autoref{fig:result:street}(bottom) corresponds to the use case in which the user is aware of the location of a specific feature. 
By choosing the reference region to be completely within the feature, focusing on similarity (low divergence) reveals the other similar features.
\autoref{fig:result:heated}(right) is a case in which the user expects turbulence and aims to declutter the visualization. 
When the reference region is chosen in such a way that turbulence is captured within the reference region, abnormal behavior is more easily identifiable.
Note that \autoref{fig:result:street}, reveals structures akin to LCS. 
We assume this to be the result of choosing a reference region with a low (or high) $\alpha$, since dissimilarity to a low (or high) alpha is equivalent to a high magnitude of stretching.

\section{Conclusions}

In this work, we provided a novel user-guided approach to visualize similarity.
Our technique utilizes the concept of strain progression to encode the strains and rotation along a pathline as distributions.
We then compare these distributions using the Jensen-Shannon divergence, which measures the difference between two distributions.
The benefit of our strain-based method is that the entire evolution of fluid parcels is considered when computing similarities.

In the future, we will expand our method to three dimensions and apply the approach to turbulent flows.
This is of interest because visualizations of turbulent flows are often cluttered, as self-similar patterns occur on many scales.
\autoref{fig:result:heated} suggests, that our method is able to "encode" turbulence in the reference distribution and potentially increasing the quality of the similarity visualization. 
We also want to improve on the description of the reference distribution and explore clustering techniques.

In addition, we will explore techniques to speed up the computation and improve the memory footprint, which is critical when taking 3D flows into account.

\acknowledgments{
This work was partially funded by the German Federal Ministry of Education and Research within the project Competence Center for Scalable Data Services and Solutions (ScaDS) Dresden/Leipzig (BMBF 01IS14014B).}

\bibliographystyle{abbrv-doi}

\bibliography{template}
\end{document}